\documentclass[11pt]{article}
\usepackage[latin9]{inputenc}
\usepackage{geometry}
\usepackage[english]{babel}
\usepackage{array}
\usepackage{float}
\usepackage{url}
\usepackage{bm}
\usepackage{multirow}
\usepackage{amsmath}
\usepackage{amsthm}
\usepackage{amssymb}
\usepackage{setspace}
\usepackage[authoryear]{natbib}
\usepackage[unicode=true,pdfusetitle,
 bookmarks=true,bookmarksnumbered=false,bookmarksopen=false,
 breaklinks=false,pdfborder={0 0 0},pdfborderstyle={},backref=false,colorlinks=false]
 {hyperref}

\makeatletter

\let\SF@@footnote\footnote
\def\footnote{\ifx\protect\@typeset@protect
    \expandafter\SF@@footnote
  \else
    \expandafter\SF@gobble@opt
  \fi
}
\expandafter\def\csname SF@gobble@opt \endcsname{\@ifnextchar[%]
  \SF@gobble@twobracket
  \@gobble
}
\edef\SF@gobble@opt{\noexpand\protect
  \expandafter\noexpand\csname SF@gobble@opt \endcsname}
\def\SF@gobble@twobracket[#1]#2{}
\providecommand{\tabularnewline}{\\}
\floatstyle{ruled}
\newfloat{algorithm}{tbp}{loa}
\providecommand{\algorithmname}{Algorithm}
\floatname{algorithm}{\protect\algorithmname}

\bibpunct{(}{)}{,}{a}{,}{,}
\usepackage{url}
\theoremstyle{plain}

\usepackage{lscape}
\usepackage[bottom]{footmisc}

\newtheorem{ass}{Assumption}

\makeatother

\begin{document}
\title{Jacobian-free Efficient Pseudo-Likelihood (EPL) Algorithm\thanks{The replication code of the numerical experiments in this article is available at \protect\url{https://github.com/takeshi-fukasawa/EPL}.}}
\author{Takeshi Fukasawa\thanks{Graduate School of Public Policy, The University of Tokyo. 7-3-1 Hongo, Tokyo, Japan. E-mail: fukasawa3431@gmail.com.}}
\maketitle
\begin{abstract}
This study proposes a simple procedure to compute Efficient Pseudo Likelihood (EPL) estimator proposed by \citet{dearing2024efficient} for estimating dynamic discrete games, without computing Jacobians of equilibrium constraints. EPL estimator is efficient, convergent, and computationally fast. However, the original algorithm requires deriving and coding the Jacobians, which are cumbersome and prone to coding mistakes especially when considering complicated models. The current study proposes to avoid the computation of Jacobians by combining the ideas of numerical derivatives (for computing Jacobian-vector products) and the Krylov method (for solving linear equations). It shows good computational performance of the proposed method by numerical experiments.


\pagebreak{}
\end{abstract}

\section{Introduction}

Structural estimations of models with equilibrium constraints are prevalent in economics. Examples include dynamic discrete games with incomplete information, and \citet{dearing2024efficient} recently proposed Efficient Pseudo Likelihood (EPL) estimator for estimating the model. The idea is to sequentially update structural parameters $\theta$ and nuisance parameters $Y$ (choice-specific value functions) by introducing Newton-like mappings. Unlike conditional choice probability estimators and Nested Pseudo Likelihood (NPL) estimators (\citealp{aguirregabiria2007sequential}), the EPL estimator is guaranteed to be efficient under regularity conditions. Moreover, MLE is a fixed point of the EPL iterations. Unlike the MPEC (Mathematical Programming with Equilibrium Constraints) approach (\citealp{egesdal2015estimating}), the algorithm does not require sophisticated software. Compared to the Nested-Fixed Point (NFXP) approach, the EPL is computationally fast and robust even when multiple equilibria exist.

One obstacle to implementing the EPL algorithm is the need to compute Jacobians of the equilibrium constraints concerning $Y$. First, analytically deriving the Jacobians is cumbersome and prone to coding mistakes, especially when considering complicated models.\footnote{For instance, \citet{igami2017estimating} introduced nonstationarity and sequential move assumptions in dynamic games.} Second, the Jacobian is an $|Y|\times|Y|$-dimensional matrix, the memory requirement and the computational cost can be high, especially when the dimension of the nuisance parameters is large.

The current study proposes a simple procedure to avoid the problem mentioned above. The idea is to combine numerical derivatives (for computing Jacobian-vector products) and the Krylov method (for solving linear equations).\footnote{Analogous idea has been used to apply Newton's method for solving nonlinear equations. See Section 3 of \citet{kelley2003solving}.} With this strategy, we can completely avoid computing the Jacobians. I show good computational performance of the proposed method by numerical experiments.

Note that the proposed method is not necessarily as fast as the original EPL algorithm. Nevertheless, the proposed method does not require computing Jacobians, and need less storage than the original EPL. Moreover,the proposed method is still much faster than the NFXP. The proposed method would be worth trying even when practitioners want to experiment whether the EPL works well, before coding complicated Jacobians. 

\section{Algorithm}

\subsection{EPL Algorithm}

Efficient Pseudo-Likelihood (EPL) estimator is designed to estimate the following problem:

\begin{eqnarray*}
\left(\widehat{\theta}_{MLE},\widehat{Y}_{MLE}\right) & = & \arg\max_{(\theta,Y)\in\Theta\times\mathcal{Y}}Q_{N}(\theta,Y)\\
 & s.t. & G(\theta,Y)=0
\end{eqnarray*}

where $Q_{N}\left(\theta,Y\right)\equiv\frac{1}{N}\sum_{i=1}^{N}\ln Pr\left(w_{i}|\theta,Y\right)$. $i$ denotes a sample, and $w_{i}$ denotes an observed outcome. $Y$ is called a nuisance parameter. In the dynamic discrete game discussed in Section \ref{sec:Numerical-experiments:-Dynamic}, choice-specific value function $v$ corresponds to $Y$. 

Here, we further assume linearity concerning parameters $\theta$: 
\[
G\left(\theta,Y\right)=H\left(Y\right)\theta+z\left(Y\right).
\]
Here, $H\left(Y\right)\equiv\left(h_{i}\left(Y\right)\right)_{i=1,\cdots,|\Theta|}$ is a $|\mathcal{Y}|\times|\Theta|$-dimensional matrix, and $z\left(Y\right)$ is a $|\mathcal{Y}|$-dimensional vector.

Algorithm \ref{alg:EPL-algorithm} shows the steps of the EPL algorithm.

\begin{algorithm}[H]
\caption{EPL algorithm\label{alg:EPL-algorithm}}

\begin{enumerate}
\item Set initial values $\widehat{\theta}_{0},\widehat{Y}_{0}$ and tolerance levels $\epsilon_{\theta},\epsilon_{Y}$
\item Iterate the following $(k=1,2,\cdots)$:
\begin{enumerate}
\item Compute $\left(\nabla_{Y}G\left(\widehat{\theta}_{k-1},\widehat{Y}_{k-1}\right)\right)^{-1}H\left(\widehat{Y}_{k-1}\right)$ and $\left(\nabla_{Y}G\left(\widehat{\theta}_{k-1},\widehat{Y}_{k-1}\right)\right)^{-1}z\left(\widehat{Y}_{k-1}\right)$
\item Compute $\widehat{\theta}_{k}=\arg\max_{\theta\in\Theta}Q_{N}\left(\theta,\Upsilon(\theta,\widehat{\gamma}_{k-1})\right)$

where $\widehat{\gamma}_{k}\equiv\left(\widehat{\theta}_{k},\widehat{Y}_{k}\right)$ and

\begin{eqnarray*}
\Upsilon(\theta,\hat{\gamma}_{k-1}) & \equiv & \widehat{Y}_{k-1}-\left(\nabla_{Y}G\left(\widehat{\theta}_{k-1},\widehat{Y}_{k-1}\right)\right)^{-1}\left(G\left(\theta,\widehat{Y}_{k-1}\right)\right)\\
 & = & \widehat{Y}_{k-1}-\left(\left(\nabla_{Y}G\left(\widehat{\theta}_{k-1},\widehat{Y}_{k-1}\right)\right)^{-1}\left(H\left(\widehat{Y}_{k-1}\right)\right)\right)\theta-\\
 &  & \left(\left(\nabla_{Y}G\left(\widehat{\theta}_{k-1},\widehat{Y}_{k-1}\right)\right)^{-1}\left(z\left(\widehat{Y}_{k-1}\right)\right)\right)
\end{eqnarray*}

\item Compute $\widehat{Y}_{k}=\Upsilon\left(\widehat{\theta}_{k},\widehat{\gamma}_{k-1}\right)$
\item If $\left\Vert \theta_{k}-\theta_{k-1}\right\Vert \leq\epsilon_{\theta}$ and $\left\Vert Y_{k}-Y_{k-1}\right\Vert \leq\epsilon_{Y}$, exit the iteration. 

Otherwise, go back to Step 2(a).
\end{enumerate}
\end{enumerate}
\end{algorithm}

\subsubsection{Obstacles to the EPL algorithm with analytical Jacobians}

In the EPL algorithm, Step 2(a) would be the largest obstacle to the implementation. In the step, we need to compute the values of $\left(\nabla_{Y}G\left(\widehat{\theta}_{k-1},\widehat{Y}_{k-1}\right)\right)^{-1}H\left(\widehat{Y}_{k-1}\right)$ and $\left(\nabla_{Y}G\left(\widehat{\theta}_{k-1},\widehat{Y}_{k-1}\right)\right)^{-1}z\left(\widehat{Y}_{k-1}\right)$. Coding $\nabla_{Y}G\left(\widehat{\theta}_{k-1},\widehat{Y}_{k-1}\right)$ is cumbersome, and prone to coding mistakes, especially for complicated models. In addition, computing $|Y|\times|Y|$-dimensional matrix $\nabla_{Y}G\left(\widehat{\theta}_{k-1},\widehat{Y}_{k-1}\right)$ requires much memory requirement and computation time when $|Y|$ is large.\footnote{We might be able to mitigate the problem by utilizing sparsity. However, the strategy does not work for dense problems.}

\subsection{Proposed procedure for Step 2(a) of the EPL algorithm}

The proposed procedure builds on the ideas of the Krylov method and numerical derivatives. We discuss them before describing the proposed method.

\subsubsection{Krylov iterative method\protect\footnote{See \citet{saad2003iterative} for details.}}

Suppose we want to solve $Ad=b$ for $d\in\mathbb{R}^{m}$, where $A\in\mathbb{R}^{m\times m}$, $b\in\mathbb{R}^{m}$. Algorithm \ref{alg:Krylov} shows the basic structure of the Krylov method.

\begin{algorithm}[H]
\caption{Basic structure of the Krylov method\label{alg:Krylov}}

\begin{itemize}
\item Goal: solve $Ad=b$ for $d\in\mathbb{R}^{m}$, where $A\in\mathbb{R}^{m\times m}$, $b\in\mathbb{R}^{m}$.
\item Input: $d_{0}$, $b$, $\left(A\ \text{or}\ g:\mathbb{R}^{m}\rightarrow\mathbb{R}^{m}\text{\ such that\ }g(\bm{v})=A\bm{v}\right)$ , and tolerance level $\epsilon_{GMRES}$
\item Output: $d^{*}$
\end{itemize}
\begin{enumerate}
\item Compute $r_{0}\equiv Ad-b$. Let $w_{0}\equiv r_{0}$.
\item Iterate the following $(n=1,2,\cdots)$:
\begin{enumerate}
\item Compute $w_{n}\equiv Aw_{n-1}$ (which is equivalent to $A^{n}r_{0}$)
\item Compute 
\begin{eqnarray*}
\gamma_{n}^{*} & \equiv & \arg\min_{\gamma\equiv\left(\gamma_{n,i}\right)_{i=0,\cdots,n-1}}\left\Vert A\left(d_{0}+\sum_{i=0}^{n-1}\gamma_{n,i}A^{i}r_{0}\right)-b\right\Vert _{2}^{2}\\
 & = & \arg\min_{\gamma\equiv\left(\gamma_{n,i}\right)_{i=0,\cdots,n-1}}\left\Vert r_{0}+\sum_{i=0}^{n-1}\gamma_{n,i}w_{i+1}\right\Vert _{2}^{2}
\end{eqnarray*}
\item If $\left\Vert r_{0}+\sum_{i=0}^{n-1}\gamma_{n,i}w_{i+1}\right\Vert \leq\epsilon_{GMRES}$, let $d^{*}=d_{0}+\sum_{i=0}^{n-1}\gamma_{n,i}^{*}A^{i}r_{0}=d_{0}+\sum_{i=0}^{n-1}\gamma_{n,i}^{*}w_{i+1}$ and exit the iteration. 

Otherwise, go back to Step 2(a).
\end{enumerate}
\end{enumerate}
\end{algorithm}

As shown in Algorithm \ref{alg:Krylov}, the matrix $A$ appears only in the form of Jacobian-vector product $A\bm{v}$, where $\bm{v}$ denotes a vector. It implies that we can run the Krylov method if a function (linear operator) $g:\mathbb{R}^{m}\rightarrow\mathbb{R}^{m}$ such that $g(\bm{v})=A\bm{v}$ is specified. GMRES algorithm, one kind of the Krylov method, is available in many programming languages (e.g., MATLAB, Scipy package in Python, IterativeSolvers package in Julia) with the feature.

\subsubsection{Numerical derivatives}

Generally, by Taylor's theorem, twice differentiable function $f$ satisfies the following equations:

\begin{eqnarray*}
f(x+\epsilon v) & = & f(x)+\left(\nabla_{x}f(x)\right)\left(\epsilon v\right)+\frac{1}{2}\left(\epsilon v\right)^{T}\nabla^{2}f(\theta)\left(\epsilon v\right)+\epsilon^{3}O\left(\left\Vert v\right\Vert ^{3}\right),\\
f(x-\epsilon v) & = & f(x)-\left(\nabla_{x}f(x)\right)\left(\epsilon v\right)+\frac{1}{2}\left(\epsilon v\right)^{T}\nabla^{2}f(\theta)\left(\epsilon v\right)+\epsilon^{3}O\left(\left\Vert v\right\Vert ^{3}\right).
\end{eqnarray*}

Hence, $\left\Vert \frac{f(x+\epsilon v)-f(x-\epsilon v)}{2\epsilon}-\left(\nabla_{x}f(x)\right)v\right\Vert \leq\epsilon^{2}O\left(\left\Vert p\right\Vert ^{3}\right)$ holds. By taking small $\epsilon$, we can approximate the value of $\left(\nabla_{x}f(x)\right)v$ by $\frac{f(x+\epsilon v)-f(x-\epsilon v)}{2\epsilon}$ well.

\subsubsection{Proposed procedure}

By combining the two ideas, we can compute $\left(\nabla_{Y}G\left(\widehat{\theta}_{k-1},\widehat{Y}_{k-1}\right)\right)^{-1}\left(H\left(\widehat{Y}_{k-1}\right)\right)$ and $\left(\nabla_{Y}G\left(\widehat{\theta}_{k-1},\widehat{Y}_{k-1}\right)\right)^{-1}\left(z\left(\widehat{Y}_{k-1}\right)\right)$ without explicitly computing $\nabla_{Y}G\left(\widehat{\theta}_{k-1},\widehat{Y}_{k-1}\right)$. Algorithm \ref{alg:Jacobian-free-EPL} shows the proposed procedure.

\begin{algorithm}[H]
\caption{Proposed procedure to compute $\left(\nabla_{Y}G\left(\widehat{\theta}_{k-1},\widehat{Y}_{k-1}\right)\right)^{-1}H\left(\widehat{Y}_{k-1}\right)$ and $\left(\nabla_{Y}G\left(\widehat{\theta}_{k-1},\widehat{Y}_{k-1}\right)\right)^{-1}z\left(\widehat{Y}_{k-1}\right)$ (Step 2(a) of Algorithm \ref{alg:EPL-algorithm})\label{alg:Jacobian-free-EPL}}

\begin{enumerate}
\item For $i=1,\cdots,|\Theta|$, 

Specify initial value $d_{0}$. Apply the GMRES algorithm by using the following inputs: 
\begin{itemize}
\item vector $d_{0}$ (initial value)
\item vector $h_{i}\left(\widehat{Y}_{k-1}\right)$ as $b$
\item function $\widetilde{g}(\bm{v})=\frac{G\left(\widehat{\theta}_{k-1},\widehat{Y}_{k-1}+\epsilon\bm{v}\right)-G\left(\widehat{\theta}_{k-1},\widehat{Y}_{k-1}-\epsilon\bm{v}\right)}{2\epsilon}$ as $g$
\end{itemize}
Let the output be $d_{i}^{H*}$.
\item Analogously, using the vector $z\left(\widehat{Y}_{k-1}\right)$ as $b$, apply the GMRES algorithm.

Let the output be $d^{z*}$.
\item $d^{H*}\equiv\left(d_{i}^{H*}\right)_{i=1,\cdots,|\Theta|}$ and $d^{z*}$ can be regarded as the value of $\left(\nabla_{Y}G\left(\widehat{\theta}_{k-1},\widehat{Y}_{k-1}\right)\right)^{-1}\left(H\left(\widehat{Y}_{k-1}\right)\right)$ and $\left(\nabla_{Y}G\left(\widehat{\theta}_{k-1},\widehat{Y}_{k-1}\right)\right)^{-1}\left(z\left(\widehat{Y}_{k-1}\right)\right)$.
\end{enumerate}
\end{algorithm}

Note that the convergence properties of the EPL algorithm do not change as long as $d^{H*}$ and $d^{z*}$ approximate $\left(\nabla_{Y}G\left(\widehat{\theta}_{k-1},\widehat{Y}_{k-1}\right)\right)^{-1}\left(H\left(\widehat{Y}_{k-1}\right)\right)$ and $\left(\nabla_{Y}G\left(\widehat{\theta}_{k-1},\widehat{Y}_{k-1}\right)\right)^{-1}\left(z\left(\widehat{Y}_{k-1}\right)\right)$ well. Concerning the numerical accuracy, see the discussion in Appendix \ref{sec:Numerical-errors}.

\section{Numerical experiments: Dynamic Discrete Game\label{sec:Numerical-experiments:-Dynamic}}

To show the performance of the proposed method, we consider a stationary dynamic discrete game of incomplete information as in \citet{aguirregabiria2007sequential}, \citet{egesdal2015estimating}, and \citet{dearing2024efficient}.

\subsection{Setting}

Time is discrete, and indexed by $t=1,2,\cdots$. In each market, there are $|\mathcal{J}|$ firms, and they are indexed by $j\in\{1,2,\cdots,|\mathcal{J}|\}$. Given observable states $x_{t}$ and private information only observed to each firm $\epsilon_{t}^{j}$, each firm simultaneously chooses its action $a_{t}^{j}\in\mathcal{A}=\{0,1,\cdots,|\mathcal{A}|-1\}$.

Agents maximize the expected discounted utility $E\left\{ \sum_{s=0}^{\infty}\beta^{s-t}\left[\overline{u}^{j}\left(x_{s},a_{s}^{j},a_{s}^{-j};\theta_{u}\right)+\epsilon_{s}^{j}(a_{s}^{j})\right]\left|x_{t},\epsilon_{t}^{j}\right.\right\} $, where $\beta\in(0,1)$ denotes the discount factor. $\theta_{u}$ and $\theta_{f}$ represent parameters concerning utility and state transitions respectively, and let $\theta\equiv(\theta_{u},\theta_{f})$. Under the assumptions of conditional independence, independent private values, finite observed state space $\left(x_{t}\in\chi=\{1,2,\cdots,|\chi|\}\right)$, and Markov perfect equilibrium, choice-specific value function $v^{j}\in\mathbb{R}^{|\chi|\times|\mathcal{A}|}$ satisfies the following equation\footnote{See Lemma 1 of \citet{dearing2024efficient} for the proof.}:

\begin{eqnarray*}
v^{j}(x,a^{j}) & = & \Phi_{v}^{j}(x,a^{j};v^{j},v^{-j},\theta)\\
 & \equiv & u^{j}\left(x,a;\Lambda^{-j}(v^{-j}),\theta_{u}\right)+\beta\sum_{x^{\prime}}f^{j}\left(x^{\prime}|x,a^{j};\Lambda^{-j}(v^{-j}),\theta_{f}\right)S\left(v^{j}(x^{\prime})\right).
\end{eqnarray*}

Here, we define a function $\Phi_{v}:\Theta\times\mathbb{R}^{|\mathcal{J}|\times|\chi|\times|\mathcal{A}|}\rightarrow\mathbb{R}^{|\mathcal{J}|\times|\chi|\times|\mathcal{A}|}$. $\Lambda^{j}(x,a^{j};v^{j})$ is the choice probability that agent $j$ chooses action $j$ in state $x$, conditional on having choice-specific value function $v^{j}$. $S(\cdot)$ is McFadden's surplus function. We also define $\Lambda^{-j}(v^{-j})\equiv\left(\Lambda^{1}(v^{1}),\cdots,\Lambda^{j-1}(v^{j-1}),\Lambda^{j+1}(v^{j+1}),\cdots,\Lambda^{|\mathcal{J}|}(v^{|\mathcal{J}|})\right)$, and:

\begin{eqnarray*}
u^{j}\left(x,a^{j},\Lambda^{-j}(v^{-j}),\theta_{u}\right) & \equiv & \sum_{a^{-j}\in\mathcal{A}^{|\mathcal{J}|-1}}\Lambda^{-j}\left(x,a^{-j};v^{-j}\right)\overline{u}\left(x,a^{j},a^{-j};\theta_{u}\right),\\
f^{j}\left(x^{\prime}|x,a^{j};\Lambda^{-j}(v^{-j});\theta_{f}\right) & \equiv & \sum_{a^{-j}\in\mathcal{A}^{|\mathcal{J}|-1}}\Lambda^{-j}\left(x,a^{-j};v^{-j}\right)f\left(x^{\prime}|x,a^{j},a^{-j};\theta_{f}\right).
\end{eqnarray*}

Choice-specific value function $v$ is a solution of a nonlinear equation $G(\theta,v)\equiv v-\Phi_{v}(\theta,v)=0$ given $\theta$.

We consider the setting where firms make entry/exit decisions $(a^{j}=1,0)$. Concerning $\overline{u}^{j}$, we assume:

\begin{eqnarray*}
\overline{u}^{j}(x_{t},a_{t}^{j},a_{t}^{-j};\theta_{u}) & = & \begin{cases}
\theta_{FC,j}+\theta_{RS}s_{t}-\theta_{RN}\ln\left(1+\sum_{l\neq j}a_{t}^{l}\right)-\theta_{EC}\left(1-a_{t-1}^{j}\right) & \text{if}\ a_{t}^{j}=1,\\
0 & \text{if}\ a_{t}^{j}=0.
\end{cases}
\end{eqnarray*}
where $x_{t}\equiv\left(s_{t},\left\{ a_{t-1}^{j}\right\} _{j\in\mathcal{J}}\right)$ and $s_{t}\in\{1,\cdots,|\mathcal{S}|\}$ denotes the market size.

Then, $\overline{u}^{j}(x_{t},a_{t}^{j},a_{t}^{-j};\theta)=\exists h(x_{t},a_{t}^{j},a_{t}^{-j})^{\prime}\theta$ holds, and $G(\theta,v)$ can be represented as $G(\theta,v)=\exists H(v)\theta+\exists z(v)$ (linearity concerning $\theta$). We further assume $\epsilon$ follows i.i.d. type-I extreme value distribution.

The experiments are conducted on a laptop computer with the CPU AMD Ryzen 5 6600H 3.30 GHz, 16.0 GB of RAM, Windows 11 64-bit, and MATLAB 2022b. The experiments are based on the code building on the replication code of \citet{dearing2024efficient}. As in \citet{dearing2024efficient}, we choose parameter values $|\mathcal{J}|=5,|\mathcal{S}|=5,\theta_{FC,j}=-2+0.1\times j,\theta_{RS}=\theta_{RN}=\theta_{EC}=1.0$. Sample size $N$ is set to 1600. Concerning $\theta_{RN}$, we use $\theta_{RN}=4$. We assume the EPL iteration converges when $||\theta^{(n+1)}-\theta^{(n)}||_{\infty}$ and $||v^{(n+1)}-v^{(n)}||_{\infty}$ are less than (1E-2)$\slash K$, where $K$ denotes the number of parameters. Concerning the initial values, we initialize k-NPL with estimated semiparametric logit choice probabilities, and then initialize 1-EPL using the consistent parameter estimates and value function from the 1-NPL iteration.

As shown in the replication code of \citet{dearing2024efficient}, we can use the following sparsity of the Jacobian $\nabla_{Y}G\left(\widehat{\theta}_{k-1},\widehat{Y}_{k-1}\right)$ when computing the matrix: $\frac{\partial\Phi^{j}(x_{t},a_{t}^{j};v^{j},v^{-j},\theta)}{\partial v^{l}\left(\widetilde{x}_{t},a_{t}^{l}\right)}=0\ \text{if}\ l\neq j\ \text{and}\ x\neq\widetilde{x}$. Though the Jacobian's total number of elements is $|\chi|^{2}|\mathcal{J}|^{2}|\mathcal{A}|^{2}$, we have to store only $|\chi|^{2}|\mathcal{J}||\mathcal{A}|^{2}+|\chi|\left(|\mathcal{J}|^{2}-|\mathcal{J}|\right)|\mathcal{A}|^{2}$ elements when utilizing the sparsity structure. In contrast, in the proposed Jacobian-free method, we only have to compute the $\left|\chi\right|\times\left|\chi\right|$-dimensional state transition matrix, which requires less memory.

To compute numerical derivatives in the proposed algorithm, I use $\epsilon=\frac{\sqrt[3]{\mathbf{u}}}{\max\left\{ ||v||_{\infty},\text{1E-8}\right\} }$, where $\mathbf{u}$ denotes the machine precision.\footnote{Generally, $\left\Vert \frac{\widetilde{f}(x+\epsilon v)-\widetilde{f}(x-\epsilon v)}{2\epsilon}-\nabla f(x)^{T}v\right\Vert \leq\epsilon^{2}O\left(\left\Vert v\right\Vert ^{3}\right)+\frac{L_{f}}{\epsilon}$ holds for twice differentiable function $f$ when $\left\Vert \widetilde{f}(x)-f(x)\right\Vert \leq L_{f}\ \forall x$. The right hand side of the inequality is minimized when $\epsilon=\frac{L_{f}}{\sqrt[3]{O\left(\left\Vert v\right\Vert ^{3}\right)}}$. In the current setting, evaluating $G(\theta,v)$ incurs rounding errors, and $L_{f}$ would be around the machine precision $\mathbf{u}$. Hence, I choose the specification $\epsilon=\frac{\sqrt[3]{\mathbf{u}}}{\max\left\{ ||v||_{\infty},\text{1E-8}\right\} }$. Note that 1E-8 is introduced to avoid the division by 0 when $||v||_{\infty}$ equals 0.} As $\epsilon_{GMRES}$ (tolerance level in the GMRES iteration), I use 1E-5.

\subsection{Results}

Table \ref{tab:EPL-results} compares the proposed EPL algorithm with the original EPL algorithm. Here, $k$-EPL denotes the EPL estimator where the number of iterations is set to $k$. $\infty$-EPL denotes the estimator after the convergence of the iterations. ``Anal.'' denotes the EPL estimator using analytical Jacobians and mldivide function (non-iterative method to solve linear equations) in MATLAB. ``Krylov'' denotes the EPL estimator using analytical Jacobians and gmres function (iterative method to solve linear equations) in MATLAB. Finally, ``JF'' denotes the EPL estimator based on the proposed Jacobian-free approach. The results show that the difference between the EPL-JF and EPL-Analytical is mostly negligible. Concerning the computation time, EPL-JF is several times slower than EPL-Analytical and EPL-Krylov. Nevertheless, considering the benefit from avoiding coding Jacobians, EPL-JF would be attractive.

Note that EPL-JF requires much less computation time, compared to the NFXP algorithm using numerical derivatives in the outer-loop. The latter is also Jacobian-free, and we denote it as NFXP-JF. Table \ref{tab:Comparison-with-NFXP} compares the performance,\footnote{Concerning the inner-loop of NFXP, I use the fixed-point iterations using the mapping $\Phi_{v}$. The tolerance level of the inner-loop iteration is set to 1E-12. To accelerate the convergence, I combine the Anderson acceleration method. Though I also experimented other fixed-point mappings and acceleration methods, the results were not largely different. Concerning the outer-loop optimization, the tolerance level is set to 1E-6, and I use fminunc function in MATLAB with central-difference numerical derivatives. In the current study, the inner-loop iterations are started from only one initial values of $v$. Generally, when we cannot rule out the possibility of multiple equilibria, we should start from several initial values. When we introduce such procedure, the computational cost of the NFXP algorithm would be much larger.} and we can see that the computational cost of EPL-JF is tens of times smaller than the NFXP. 

\begin{table}[H]
\caption{Performance of the proposed EPL algorithm\label{tab:EPL-results}}

\renewcommand{\arraystretch}{0.8}
\begin{centering}
{\footnotesize{}}%
\begin{tabular}{cccccccccccccc}
\hline 
 &  & \multicolumn{3}{c}{{\tiny{}1-EPL}} & \multicolumn{3}{c}{{\tiny{}2-EPL}} & \multicolumn{3}{c}{{\tiny{}3-EPL}} & \multicolumn{3}{c}{{\tiny{}$\infty$-EPL}}\tabularnewline
 &  & {\tiny{}Anal.} & {\tiny{}Krylov} & {\tiny{}JF} & {\tiny{}Anal.} & {\tiny{}Krylov} & {\tiny{}JF} & {\tiny{}Anal.} & {\tiny{}Krylov} & {\tiny{}JF} & {\tiny{}Anal.} & {\tiny{}Krylov} & {\tiny{}JF}\tabularnewline
\hline 
\hline 
\multirow{2}{*}{{\tiny{}$\log_{10}(\text{Difference\ in\ }\theta)$}} & {\tiny{}Mean} & {\tiny{}-} & {\tiny{}-6.328} & {\tiny{}-6.323} & {\tiny{}-} & {\tiny{}-6.072} & {\tiny{}-6.066} & {\tiny{}-} & {\tiny{}-6.032} & {\tiny{}-6.024} & {\tiny{}-} & {\tiny{}-6.034} & {\tiny{}-6.03}\tabularnewline
 & {\tiny{}Max} & {\tiny{}-} & {\tiny{}-5.091} & {\tiny{}-5.112} & {\tiny{}-} & {\tiny{}-4.872} & {\tiny{}-4.873} & {\tiny{}-} & {\tiny{}-4.864} & {\tiny{}-4.873} & {\tiny{}-} & {\tiny{}-4.744} & {\tiny{}-4.743}\tabularnewline
\hline 
\multirow{3}{*}{{\tiny{}Iterations}} & {\tiny{}Median} & {\tiny{}1} &  & {\tiny{}1} & {\tiny{}2} &  & {\tiny{}2} & {\tiny{}3} &  & {\tiny{}3} & {\tiny{}5} & {\tiny{}5} & {\tiny{}5}\tabularnewline
 & {\tiny{}Max} & {\tiny{}1} &  & {\tiny{}1} & {\tiny{}2} &  & {\tiny{}2} & {\tiny{}3} &  & {\tiny{}3} & {\tiny{}8} & {\tiny{}8} & {\tiny{}8}\tabularnewline
 & {\tiny{}Non-Conv.} &  &  &  &  &  &  &  &  &  & {\tiny{}0\%} & {\tiny{}0\%} & {\tiny{}0\%}\tabularnewline
\hline 
\multirow{4}{*}{{\tiny{}Time(secs)}} & {\tiny{}Total} & {\tiny{}13.377} & {\tiny{}6.914} & {\tiny{}45.273} & {\tiny{}25.254} & {\tiny{}16.793} & {\tiny{}147.737} & {\tiny{}37.026} & {\tiny{}26.111} & {\tiny{}247.475} & {\tiny{}65.822} & {\tiny{}49.389} & {\tiny{}506.214}\tabularnewline
 & {\tiny{}Mean} & {\tiny{}0.134} & {\tiny{}0.069} & {\tiny{}0.453} & {\tiny{}0.253} & {\tiny{}0.168} & {\tiny{}1.477} & {\tiny{}0.37} & {\tiny{}0.261} & {\tiny{}2.475} & {\tiny{}0.658} & {\tiny{}0.494} & {\tiny{}5.062}\tabularnewline
 & {\tiny{}Median} & {\tiny{}0.14} & {\tiny{}0.049} & {\tiny{}0.341} & {\tiny{}0.264} & {\tiny{}0.129} & {\tiny{}1.257} & {\tiny{}0.391} & {\tiny{}0.208} & {\tiny{}2.033} & {\tiny{}0.652} & {\tiny{}0.436} & {\tiny{}4.589}\tabularnewline
 & {\tiny{}Med/Iter} & {\tiny{}0.134} & {\tiny{}0.069} & {\tiny{}0.453} & {\tiny{}0.126} & {\tiny{}0.084} & {\tiny{}0.739} & {\tiny{}0.123} & {\tiny{}0.087} & {\tiny{}0.825} & {\tiny{}0.12} & {\tiny{}0.089} & {\tiny{}0.917}\tabularnewline
\hline 
\end{tabular}{\footnotesize\par}
\par\end{centering}
\renewcommand{\arraystretch}{1}

{\footnotesize{}Note. Based on 100 simulations.}{\footnotesize\par}
\end{table}

\begin{table}[H]
\caption{Comparison with NFXP-JF\label{tab:Comparison-with-NFXP}}

\begin{centering}
{\footnotesize{}}%
\begin{tabular}{cccc}
\hline 
 &  & {\footnotesize{}$\infty$-EPL-JF} & {\footnotesize{}NFXP-JF}\tabularnewline
\hline 
\hline 
\multirow{2}{*}{{\footnotesize{}$\log_{10}(\text{Difference\ in\ }\theta)$}} & {\footnotesize{}Mean} & \multicolumn{2}{c}{{\footnotesize{}-4.261}}\tabularnewline
 & {\footnotesize{}Max} & \multicolumn{2}{c}{{\footnotesize{}-3.345}}\tabularnewline
\hline 
\multirow{2}{*}{{\footnotesize{}Time(secs)}} & {\footnotesize{}Mean} & {\footnotesize{}2.159} & {\footnotesize{}40.948}\tabularnewline
 & {\footnotesize{}Std} & {\footnotesize{}0.788} & {\footnotesize{}13.455}\tabularnewline
\hline 
\end{tabular}{\footnotesize\par}
\par\end{centering}
{\footnotesize{}Note. Based on 3 simulations.}{\footnotesize\par}
\end{table}

\appendix

\section{Numerical errors\label{sec:Numerical-errors}}

To consider the numerical error associated with the use of the idea of numerical derivatives in the EPL algorithm, consider the numerical error of $\nabla_{Y}G\left(\widehat{\theta}_{k-1},\widehat{Y}_{k-1}\right)^{-1}b$.

Here, let $d^{*}=\nabla_{Y}G\left(\widehat{\theta}_{k-1},\widehat{Y}_{k-1}\right)^{-1}b$ be the true solution. $d^{*}$ satisfies $\left(\nabla_{Y}G\left(\widehat{\theta}_{k-1},\widehat{Y}_{k-1}\right)\right)d^{*}=b$.

Let $\widetilde{d}$ be the value obtained through the use of numerical derivatives and the GMRES algorithm, and let $JVP\left(G\left(\widehat{\theta}_{k-1},Y\right),\widehat{Y}_{k-1},d\right)$ be the value corresponding to $\left(\nabla_{Y}G\left(\widehat{\theta}_{k-1},\widehat{Y}_{k-1}\right)\right)d$ using numerical derivatives. Suppose that the norm of the numerical error concerning the computation of Jacobian-vector product is bounded by $\epsilon_{JVP}$:

\begin{eqnarray*}
\left\Vert JVP\left(G\left(\widehat{\theta}_{k-1},Y\right),\widehat{Y}_{k-1},d\right)-\left(\nabla_{Y}G\left(\widehat{\theta}_{k-1},\widehat{Y}_{k-1}\right)\right)d\right\Vert  & \leq & \epsilon_{JVP}.
\end{eqnarray*}

In addition, suppose that the GMRES iteration is terminated when 
\begin{eqnarray*}
\left\Vert JVP\left(G\left(\widehat{\theta}_{k-1},Y\right),\widehat{Y}_{k-1},d\right)-b\right\Vert  & \leq & \epsilon_{GMRES}.
\end{eqnarray*}

Then, $\widetilde{d}$ satisfies:

\begin{eqnarray*}
\left\Vert \left(\nabla_{Y}G\left(\widehat{\theta}_{k-1},\widehat{Y}_{k-1}\right)\right)\widetilde{d}-b\right\Vert  & \leq & \left\Vert JVP\left(G\left(\widehat{\theta}_{k-1},Y\right),\widehat{Y}_{k-1},\widetilde{d}\right)-\left(\nabla_{Y}G\left(\widehat{\theta}_{k-1},\widehat{Y}_{k-1}\right)\right)\widetilde{d}\right\Vert +\\
 &  & \left\Vert JVP\left(G\left(\widehat{\theta}_{k-1},Y\right),\widehat{Y}_{k-1},\widetilde{d}\right)-b\right\Vert \\
 & \leq & \epsilon_{JVP}+\epsilon_{GMRES}
\end{eqnarray*}

It implies:\footnote{See Section 3.5 of \citet{judd1998numerical} for discussions on the error bounds of the solution of linear equations.}

$\frac{\left\Vert d^{*}-\widetilde{d}\right\Vert }{\left\Vert d^{*}\right\Vert }\leq cond\left(\nabla_{Y}G\left(\widehat{\theta}_{k-1},\widehat{Y}_{k-1}\right)\right)\frac{\epsilon_{JVP}+\epsilon_{GMRES}}{\left\Vert b\right\Vert }$, where $cond(\cdot)$ denotes the condition number of a matrix.

\bibliographystyle{apalike}
\bibliography{literature}

\end{document}